\documentclass[preprint,showpacs,amsmath,amssymb,prd,nofootinbib]{revtex4}
\usepackage{epsf}
\usepackage{dcolumn}
\usepackage{amsmath}
\usepackage{amssymb}
\usepackage{graphicx}

\usepackage[breaklinks=true,colorlinks=true]{hyperref}

\begin{document}

\title{Extended real Clifford - Dirac algebra and bosonic symmetries of the Dirac equation with nonzero mass}

\author{V.M. Simulik and I.Yu. Krivsky\footnote{Email: vsimulik@gmail.com}}

\affiliation{Institute of Electron Physics of Ukrainian National Academy of Sciences,
Uzhgorod 88017, Ukraine}


\begin{abstract}
Contemporary presentation of the version 1 demonstrates briefly the development of our investigations and our future goals. The improved free of difficulties in  interpretation and printing errors version is presented. The 256-dimensional gamma matrix representations of the Clifford algebra and corresponding representations of SO(10) and SO(1,9) algebras over the field of real numbers are introduced. This turned out to be possible in the space of eight-component spinors, and not for the ordinary four-component Dirac spinors. The proposed mathematical objects allow the generalization of our results, obtained earlier for the standard Dirac equation, for the equations of higher spin and, especially, for the equations, describing the particles with spin 3/2.
\end{abstract}


\maketitle


\section{Introduction}

The start of these investigations was given in the space of standard 4-component Dirac spinors (version 1). The gamma matrix representation of 64-dimensional Clifford algebra over the field of real numbers and corresponding representation of SO(8) Lie algebra for 4-component spinors were introduced in \cite{Real-1,Real-2,Real-3,Real-4} (the review and final description is given in \cite{Book}). The realizations $\textit{C}\ell^{\mathbb{R}}$(4,2) and $\textit{C}\ell^{\mathbb{R}}$(0,6) have been considered. The role of matrix representations of such algebras in the quantum field theory was investigated in \cite{Real-1,Real-2,Real-3,Real-4,Book} as well. The example of the standard Dirac equation has been considered. The mathematical foundations for our algebraic considerations were taken from \cite{Loun,Shir,Eli,Wyb}. Therefore, in \cite{Real-1,Real-2,Real-3,Real-4,Book} and here below we have continued our 25 years period of study, in which different useful in mathematical physics representations of the Clifford--Dirac algebra in the space of 4-component functions, see, e.g., \cite{Mex,Elect}, were introduced.  

It is evident that in the space of 8-component spinors one can find much more wide (extended) representations of these algebras. Below the 256-dimensional representations $\textit{C}\ell^{\mathbb{R}}$(0,8) and $\textit{C}\ell^{\mathbb{R}}$(1,7) of the Clifford algebra together with corresponding representations of SO(10) and SO(1,9) algebras are found. Such new representations of these algebras over the field of real numbers will be useful for the additional investigations of the first order 8-component partial differential equations of the quantum field theory, see, e.g., \cite{Kri-1,Kri-2}, especially for the equation without redundant components for the spin 3/2 particle, suggested recently in \cite{Sim-1,Sim-2} and \cite{Book}.  

\section{Representation of the $\textit{C}\ell^{\mathbb{R}}$(0,8) Clifford algebra}\label{Exact}

Consider the set of nine $8\times 8$ $\Gamma$ matrices
$$\Gamma^{j}
 = \left|
\begin{array}{cccc}
0 & 0 & 0 & \sigma^{j}\\
0 & 0 & \sigma^{j} & 0\\
0 & \sigma^{j} & 0 & 0\\
\sigma^{j} & 0 & 0 & 0\\
\end{array} \right|, \, j=1,2,3, \quad \Gamma^{4}=i\left|
\begin{array}{cccc}
0 & 0 & 0 & -\mathrm{I}_{2}\\
0 & 0 & \mathrm{I}_{2} & 0\\
0 & -\mathrm{I}_{2} & 0 & 0\\
\mathrm{I}_{2} & 0 & 0 & 0\\
\end{array} \right|, \quad \Gamma^{5}=i\left|
\begin{array}{cccc}
0 & 0 & -\mathrm{I}_{2} & 0\\
0 & 0 & 0 & -\mathrm{I}_{2}\\
\mathrm{I}_{2} & 0 & 0 & 0\\
0 & \mathrm{I}_{2} & 0 & 0\\
\end{array} \right|,
 $$
\begin{equation}
\label{Main}
\Gamma^{6}=\left|
\begin{array}{cccc}
0 & 0 & \mathrm{I}_{2} & 0\\
0 & 0 & 0 & -\mathrm{I}_{2}\\
\mathrm{I}_{2} & 0 & 0 & 0\\
0 & -\mathrm{I}_{2} & 0 & 0\\
\end{array} \right|, \quad \Gamma^{7}=\left|
\begin{array}{cccc}
\mathrm{I}_{2} & 0 & 0 & 0\\
0 & \mathrm{I}_{2} & 0 & 0\\
0 & 0 & -\mathrm{I}_{2} & 0\\
0 & 0 & 0 & -\mathrm{I}_{2}\\
\end{array} \right|, \quad \mathrm{I}_{2}=\left| {{\begin{array}{*{20}c}
 1 \hfill &  0 \hfill\\
 0 \hfill & 1  \hfill\\
\end{array} }} \right|,
\end{equation}
$$\Gamma^{8}=\left|
\begin{array}{cccc}
0 & 0 & -\sigma^{2}\hat{C} & 0\\
0 & 0 & 0 & \sigma^{2}\hat{C}\\
\sigma^{2}\hat{C} & 0 & 0 & 0\\
0 & -\sigma^{2}\hat{C} & 0 & 0\\
\end{array} \right|, \quad \Gamma^{9}=\left|
\begin{array}{cccc}
0 & 0 & -i\sigma^{2}\hat{C} & 0\\
0 & 0 & 0 & i\sigma^{2}\hat{C}\\
i\sigma^{2}\hat{C} & 0 & 0 & 0\\
0 & -i\sigma^{2}\hat{C} & 0 & 0\\
\end{array} \right|,$$
where 
\begin{equation}
\label{Pauli}
\sigma^{1}=\left| {{\begin{array}{*{20}c}
 0 \hfill &  1 \hfill\\
 1 \hfill & 0  \hfill\\
 \end{array} }} \right|, \quad \sigma^{2}=\left| {{\begin{array}{*{20}c}
 0 \hfill &  -i \hfill\\
 i \hfill &   0  \hfill\\
 \end{array} }} \right|, \quad \sigma^{3}=\left| {{\begin{array}{*{20}c}
 1 \hfill &  0 \hfill\\
 0 \hfill & -1  \hfill\\
\end{array} }} \right|, \quad \sigma^{1}\sigma^{2}=i\sigma^{3}, \quad 123!-\mathrm{circle},
\end{equation}
are the standard Pauli matrices and $\hat{C}$ is the operator of complex conjugation, $\hat{C}\psi=\psi^{*}$ (the operator of involution in the Hilbert space $\mathrm{H}^{3,2}$). The operators (\ref{Main}) satisfy the anti-commutation relations 
\begin{equation}
\label{Cliff}
\Gamma ^\mathrm{A} \Gamma ^\mathrm{B} + \Gamma
^\mathrm{B}\Gamma ^\mathrm{A} = 2\delta^{\mathrm{A}\mathrm{B}},\quad \mathrm{A},\mathrm{B}=\overline{1,9},
\end{equation}
of the Clifford algebra. Nevertheless, only eight operators from the set (\ref{Main}) are linearly independent and determine the set of of the generators of the corresponding Clifford algebra. For example, such linear dependence can be demonstrated in explicit form as operator products
\begin{equation}
\label{Cond}
\Gamma^{7}=-i\Gamma^{1}\Gamma^{2}\Gamma^{3}\Gamma^{4}\Gamma^{5}\Gamma^{6},
\end{equation}
together with the following relationship $\Gamma^{1}\Gamma^{2}\Gamma^{3}\Gamma^{4}\Gamma^{5}\Gamma^{6}\Gamma^{7}\Gamma^{8}\Gamma^{9}=\mathrm{I}_{8}$. Therefore, the dimension of this algebra is $2^{8}=256$. Thus, we deal with representation of $\textit{C}\ell^{\mathbb{R}}$(0,8).

Indeed, first six operators from (\ref{Main}) form the representation of the $\textit{C}\ell^{\mathbb{C}}$(0,6) algebra (here we deal with $2^{6}=64$ dimensions). Note that such algebra is an analogue of the ordinary 16-dimensional Clifford--Dirac algebra $\textit{C}\ell^{\mathbb{C}}$(0,4) in the space of 8-component spinors. Therefore, similarly to \cite{Real-1,Real-2,Real-3,Real-4,Book} we can recalculate the complete set of 256 elements of the $\textit{C}\ell^{\mathbb{R}}$(0,8) algebra representation in the form
\begin{equation}
\label{Calc} \left\{ (64\textit{C}\ell^{\mathbb{C}}(0,6))\cup i\cdot(64\textit{C}\ell^{\mathbb{C}}(0,6))\cup \hat{C}\cdot(64\textit{C}\ell^{\mathbb{C}}(0,6))\cup i\hat{C}\cdot(64\textit{C}\ell^{\mathbb{C}}(0,6)) \right\},
\end{equation}
where $\widetilde{C}$ is the operator of complex conjugation in the space of 8-component spinors, $\widetilde{C}\psi=\psi^{*}$ (the operator of involution in the Hilbert space $\mathrm{H}^{3,8}$).

It is useful to present these 64 and 256 elements visually in explicit form, as one can found, e. g., in \cite{Good} for the standard representation of $\textit{C}\ell^{\mathbb{C}}$(0,4) of the Clifford--Dirac algebra. The corresponding formalism are given below in Appendices 1 and 2 respectively.

\section{Representation of the $\textit{C}\ell^{\mathbb{R}}$(1,7) Clifford algebra}\label{Exact-2}

Let us mark the first difference between the consideration in four-component \cite{Real-1,Real-2,Real-3,Real-4} and eight-component formalism. In the space of 4-component spinors it was impossible to introduce the representation $\textit{C}\ell^{\mathbb{R}}$(1,5) instead (or together with) $\textit{C}\ell^{\mathbb{R}}$(0,6). Here such variant is possible and is under consideration due to the more wide properties of the algebra $\textit{C}\ell^{\mathbb{R}}$(1,7):
$$\quad \Gamma^{0}=\left|
\begin{array}{cccc}
\mathrm{I}_{2} & 0 & 0 & 0\\
0 & \mathrm{I}_{2} & 0 & 0\\
0 & 0 & -\mathrm{I}_{2} & 0\\
0 & 0 & 0 & -\mathrm{I}_{2}\\
\end{array} \right|, \quad \Gamma^{j}
 = \left|
\begin{array}{cccc}
0 & 0 & 0 & \sigma^{j}\\
0 & 0 & \sigma^{j} & 0\\
0 & -\sigma^{j} & 0 & 0\\
-\sigma^{j} & 0 & 0 & 0\\
\end{array} \right|, \quad \Gamma^{4}=i\left|
\begin{array}{cccc}
0 & 0 & 0 & -\mathrm{I}_{2}\\
0 & 0 & \mathrm{I}_{2} & 0\\
0 & \mathrm{I}_{2} & 0 & 0\\
-\mathrm{I}_{2} & 0 & 0 & 0\\
\end{array} \right|, $$
\begin{equation}
\label{Main-2}
\Gamma^{5}=i\left|
\begin{array}{cccc}
0 & 0 & -\mathrm{I}_{2} & 0\\
0 & 0 & 0 & -\mathrm{I}_{2}\\
-\mathrm{I}_{2} & 0 & 0 & 0\\
0 & -\mathrm{I}_{2} & 0 & 0\\
\end{array} \right|, \quad \Gamma^{6}=\left|
\begin{array}{cccc}
0 & 0 & \mathrm{I}_{2} & 0\\
0 & 0 & 0 & -\mathrm{I}_{2}\\
-\mathrm{I}_{2} & 0 & 0 & 0\\
0 & \mathrm{I}_{2} & 0 & 0\\
\end{array} \right|,
\end{equation}
$$\Gamma^{7}=\left|
\begin{array}{cccc}
0 & 0 & \sigma^{2}\hat{C} & 0\\
0 & 0 & 0 & -\sigma^{2}\hat{C}\\
\sigma^{2}\hat{C} & 0 & 0 & 0\\
0 & -\sigma^{2}\hat{C} & 0 & 0\\
\end{array} \right|, \quad \Gamma^{8}=\left|
\begin{array}{cccc}
0 & 0 & i\sigma^{2}\hat{C} & 0\\
0 & 0 & 0 & -i\sigma^{2}\hat{C}\\
i\sigma^{2}\hat{C} & 0 & 0 & 0\\
0 & -i\sigma^{2}\hat{C} & 0 & 0\\
\end{array} \right|.$$
The operators (\ref{Main-2}) satisfy the anti-commutation relations 
\begin{equation}
\label{Cliff-2}
\Gamma ^\mathrm{\tilde{A}} \Gamma ^\mathrm{\tilde{B}} + \Gamma
^\mathrm{\tilde{B}}\Gamma ^\mathrm{\tilde{A}} = 2g^{\mathrm{\tilde{A}}\mathrm{\tilde{B}}}, \quad g=(+--------), \quad \mathrm{\tilde{A}},\mathrm{\tilde{B}}=\overline{0,8},
\end{equation}
of the Clifford algebra generators. Again, only eight operators from the set (\ref{Main-2}) are linearly independent. For example, such linear dependence can be demonstrated in explicit form as operator product $\Gamma^{0}\Gamma^{1}\Gamma^{2}\Gamma^{3}\Gamma^{4}\Gamma^{5}\Gamma^{6}=-i\mathrm{I}_{8}$ together with $\Gamma^{0}\Gamma^{1}\Gamma^{2}\Gamma^{3}\Gamma^{4}\Gamma^{5}\Gamma^{6}\Gamma^{7}\Gamma^{8}=\mathrm{I}_{8}$. Therefore, we deal with 256-dimensional representation of $\textit{C}\ell^{\mathbb{R}}$(1,7).

Further description of this algebra is similar to the given in Sec 2 above.

\section{Gamma matrix representation of SO(10) algebra}\label{Good}

Operators (\ref{Main}) generate also the 45 matrices 
\begin{equation}
\label{Spin-1}
s^{\widetilde{\mathrm{A}}\widetilde{\mathrm{B}}}=\{s^{\mathrm{A}\mathrm{B}}=-\frac{1}{4}[\Gamma
^\mathrm{A},\Gamma
^\mathrm{B}],\,s^{\mathrm{A}10}=-s^{10\mathrm{A}}=-\frac{1}{2}\Gamma
^\mathrm{A}\},\quad \widetilde{\mathrm{A}},\widetilde{\mathrm{B}}=\overline{1,10}, \quad \mathrm{A},\mathrm{B}=\overline{1,9},
\end{equation}
which satisfy the commutation relations of the generators of the Lie algebra SO(10):
\begin{equation}
\label{Comm}
[s^{\widetilde{\mathrm{A}}\widetilde{\mathrm{B}}},s^{\widetilde{\mathrm{C}}\widetilde{\mathrm{D}}}]=
\delta^{\widetilde{\mathrm{A}}\widetilde{\mathrm{C}}}s^{\widetilde{\mathrm{B}}\widetilde{\mathrm{D}}}
+\delta^{\widetilde{\mathrm{C}}\widetilde{\mathrm{B}}}s^{\widetilde{\mathrm{D}}\widetilde{\mathrm{A}}}
+\delta^{\widetilde{\mathrm{B}}\widetilde{\mathrm{D}}}s^{\widetilde{\mathrm{A}}\widetilde{\mathrm{C}}}
+\delta^{\widetilde{\mathrm{D}}\widetilde{\mathrm{A}}}s^{\widetilde{\mathrm{C}}\widetilde{\mathrm{B}}}.
\end{equation}
Note that here (as in \cite{Real-1,Real-2,Real-3,Real-4}) the anti-Hermitian realization of the SO(10) operators is chosen, for the reasons see, e.g., \cite{Real-1,Real-2,Real-3,Real-4,Book} and \cite{Eli,Wyb}. We appeal to the anti-Hermitian realizations of the generators starting from \cite{Mex,Elect}.

The explicit form of the 45 elements of the $\Gamma$ matrix representation of the SO(10) algebra is given in the Table 1.
$$\mathrm{\textbf{Table 1.} \, The \, 45 \, \, elements \, of \, the \, \Gamma \, \, matrix \,\, representation \, of \, the \, \, SO(10) \, algebra}$$
\begin{center}
\begin{tabular}{|c|c|c|c|c|c|c|c|c|c|}
\hline
\rule{0pt}{5mm} $-\frac{1}{2}\Gamma^{1}\Gamma^{2}$  & $-\frac{1}{2}\Gamma^{1}\Gamma^{3}$ & $-\frac{1}{2}\Gamma^{1}\Gamma^{4}$ & $-\frac{1}{2}\Gamma^{1}\Gamma^{5}$ & $-\frac{1}{2}\Gamma^{1}\Gamma^{6}$ & $-\frac{1}{2}\Gamma^{1}\Gamma^{7}$ & $-\frac{1}{2}\Gamma^{1}\Gamma^{8}$ & $-\frac{1}{2}\Gamma^{1}\Gamma^{9}$ & $s^{1 10}\equiv -\frac{1}{2}\Gamma^{1}$ \\
\hline
\rule{0pt}{5mm} & $-\frac{1}{2}\Gamma^{2}\Gamma^{3}$ & $-\frac{1}{2}\Gamma^{2}\Gamma^{4}$  & $-\frac{1}{2}\Gamma^{2}\Gamma^{5}$ & $-\frac{1}{2}\Gamma^{2}\Gamma^{6}$ & $-\frac{1}{2}\Gamma^{2}\Gamma^{7}$   & $-\frac{1}{2}\Gamma^{2}\Gamma^{8}$ & $-\frac{1}{2}\Gamma^{2}\Gamma^{9}$ & $s^{2 10}\equiv -\frac{1}{2}\Gamma^{2}$ \\
\hline
\rule{0pt}{5mm} &  & $-\frac{1}{2}\Gamma^{3}\Gamma^{4}$  & $-\frac{1}{2}\Gamma^{3}\Gamma^{5}$ & $-\frac{1}{2}\Gamma^{3}\Gamma^{6}$ & $-\frac{1}{2}\Gamma^{3}\Gamma^{7}$ & $-\frac{1}{2}\Gamma^{3}\Gamma^{8}$ & $-\frac{1}{2}\Gamma^{3}\Gamma^{9}$ & $s^{3 10}\equiv -\frac{1}{2}\Gamma^{3}$  \\
\hline
\rule{0pt}{5mm}  &  &  & $-\frac{1}{2}\Gamma^{4}\Gamma^{5}$ & $-\frac{1}{2}\Gamma^{4}\Gamma^{6}$ & $-\frac{1}{2}\Gamma^{4}\Gamma^{7}$ & $-\frac{1}{2}\Gamma^{4}\Gamma^{8}$ & $-\frac{1}{2}\Gamma^{4}\Gamma^{9}$ & $s^{4 10}\equiv -\frac{1}{2}\Gamma^{4}$ \\
\hline
\rule{0pt}{5mm} &  &   &  & $-\frac{1}{2}\Gamma^{5}\Gamma^{6}$ & $-\frac{1}{2}\Gamma^{5}\Gamma^{7}$ & $-\frac{1}{2}\Gamma^{5}\Gamma^{8}$ & $-\frac{1}{2}\Gamma^{5}\Gamma^{9}$ & $s^{5 10}\equiv -\frac{1}{2}\Gamma^{5}$ \\
\hline
\rule{0pt}{5mm} &  &  &  &  & $-\frac{1}{2}\Gamma^{6}\Gamma^{7}$ & $-\frac{1}{2}\Gamma^{6}\Gamma^{8}$ & $-\frac{1}{2}\Gamma^{6}\Gamma^{9}$ & $s^{6 10}\equiv -\frac{1}{2}\Gamma^{6}$ \\
\hline
\rule{0pt}{5mm} &  &  &  &  &  &  $-\frac{1}{2}\Gamma^{7}\Gamma^{8}$ &  $-\frac{1}{2}\Gamma^{7}\Gamma^{9}$ & $s^{7 10}\equiv -\frac{1}{2}\Gamma^{7}$ \\
\hline
\rule{0pt}{5mm} &  &  &  &  &  &  &  $-\frac{1}{2}\Gamma^{8}\Gamma^{9}$ & $s^{8 10}\equiv -\frac{1}{2}\Gamma^{8}$ \\
\hline
\rule{0pt}{5mm} &  &  &  &  &  &  &  &  $s^{9 10}\equiv -\frac{1}{2}\Gamma^{9}$ \\
\hline
\end{tabular}
\label{Table_1}
\end{center}

The gamma matrices in Table 1 are taken from the set (\ref{Main}).

The dimension of the SO(n) algebra is given by $\frac{n(n-1)}{2}=\frac{10\cdot 9}{2}=45$. Therefore, here we deal with SO(10) algebra representation.

Table 1 demonstrates not only the explicit form of the generators (\ref{Spin-1}) but the commutation relations (\ref{Comm}) as well. Indeed, it is evident that generators with different indices commute between each other. Further, it is evident that here we have three independent sets of SU(2) generators ($s^{1}=s^{23},\,s^{2}=s^{31},\,s^{3}=s^{12}$), which commute between each other. They are given by the following operators from the Table 1: ($-\frac{1}{2}\Gamma^{2}\Gamma^{3},-\frac{1}{2}\Gamma^{3}\Gamma^{1},-\frac{1}{2}\Gamma^{1}\Gamma^{2}$), ($-\frac{1}{2}\Gamma^{5}\Gamma^{6},-\frac{1}{2}\Gamma^{6}\Gamma^{4},-\frac{1}{2}\Gamma^{4}\Gamma^{5}$), ($-\frac{1}{2}\Gamma^{8}\Gamma^{9},-\frac{1}{2}\Gamma^{9}\Gamma^{7},-\frac{1}{2}\Gamma^{7}\Gamma^{8}$). All above mention sets of SU92) generators commute with the operator of the Foldy--Wouthuysen equation in anti-Hermitian form
\begin{equation}
\label{Foldy}
(\partial _0 +i\Gamma^{0}\sqrt{-\Delta + m^{2}} -\frac{e^{2}}{\left|\vec{x}\right|})\phi (x) = 0; \quad  x\in \mathrm{M}(1,3), \; \phi\in \left\{\mathrm{S}^{3,8}\subset\mathrm{H}^{3,8}\subset\mathrm{S}^{3,8*}\right\}.
\end{equation}
Note that in \cite{Real-1,Real-2,Real-3,Real-4,Book} in the representation of SO(8) for 4-component spinors we have only two independent SU(2) sets, which combinations make us possible to prove the Bose symmetries of the Dirac equation. Here, similarly, due to the presence of the triplet of SU(2) sets the spin 3/2 Lorentz and Poincar\'e symmetries for the equation suggested in \cite{Sim-1,Sim-2} and \cite{Book} can be found. Of course, the Bose symmetries of the 8-component Dirac equation can be found as well.     

\section{Gamma matrix representation of SO(1,9) algebra}\label{Next}

The explicit form of corresponding generators follows from the set (\ref{Main-2}). The 45 gamma matrix generators of SO(1,9) algebra are given by  
\begin{equation}
\label{Spin-2}
s^{\widehat{\mathrm{A}}\widehat{\mathrm{B}}}=\{s^{\tilde{\mathrm{A}}\tilde{\mathrm{B}}}=-\frac{1}{4}[\Gamma
^{\tilde{\mathrm{A}}},\Gamma
^{\tilde{\mathrm{B}}}],\,s^{\tilde{\mathrm{A}}9}=-s^{9\tilde{\mathrm{A}}}=\frac{1}{2}\Gamma
^{\tilde{\mathrm{A}}}\}, \quad \widehat{\mathrm{A}},\widehat{\mathrm{B}}=\overline{0,9}, \quad \mathrm{\tilde{A}},\mathrm{\tilde{B}}=\overline{0,8}. 
\end{equation}
Operators (\ref{Spin-2}) satisfy the commutation relations of the generators of the Lie algebra SO(1,9)
\begin{equation}
\label{Comm-2}
[s^{\widehat{\mathrm{A}}\widehat{\mathrm{B}}},s^{\widehat{\mathrm{C}}\widehat{\mathrm{D}}}]=
-g^{\widehat{\mathrm{A}}\widehat{\mathrm{C}}}s^{\widehat{\mathrm{B}}\widehat{\mathrm{D}}}
-g^{\widehat{\mathrm{C}}\widehat{\mathrm{B}}}s^{\widehat{\mathrm{D}}\widehat{\mathrm{A}}}
-g^{\widehat{\mathrm{B}}\widehat{\mathrm{D}}}s^{\widehat{\mathrm{A}}\widehat{\mathrm{C}}}
-g^{\widehat{\mathrm{D}}\widehat{\mathrm{A}}}s^{\widehat{\mathrm{C}}\widehat{\mathrm{B}}},
\end{equation}
where the metric tensor $g$ is already given in (\ref{Cliff-2}).

Note that here as in Sec. 4 for the same reasons the anti-Hermitian realization of the SO(1,9) operators is chosen.

The explicit form of the 45 elements of the $\Gamma$ matrix representation of the SO(1,9) algebra is given in the Table 2.

\vspace{1,7in}
 
$$\mathrm{\textbf{Table 2.} \, The \, 45 \, \, elements \, of \, the \, \Gamma \, \, matrix \,\, representation \, of \, the \, \, SO(1,9) \, algebra}$$
\begin{center}
\begin{tabular}{|c|c|c|c|c|c|c|c|c|c|}
\hline
\rule{0pt}{5mm} $\frac{1}{2}\Gamma^{0}\Gamma^{1}$  & $\frac{1}{2}\Gamma^{0}\Gamma^{2}$ & $\frac{1}{2}\Gamma^{0}\Gamma^{3}$ & $\frac{1}{2}\Gamma^{0}\Gamma^{4}$ & $\frac{1}{2}\Gamma^{0}\Gamma^{5}$ & $\frac{1}{2}\Gamma^{0}\Gamma^{6}$ & $\frac{1}{2}\Gamma^{0}\Gamma^{7}$ & $\frac{1}{2}\Gamma^{0}\Gamma^{8}$ & $s^{0 9}\equiv \frac{1}{2}\Gamma^{0}$ \\
\hline
\rule{0pt}{5mm} & $\frac{1}{2}\Gamma^{1}\Gamma^{2}$ & $\frac{1}{2}\Gamma^{1}\Gamma^{3}$  & $\frac{1}{2}\Gamma^{1}\Gamma^{4}$ & $\frac{1}{2}\Gamma^{1}\Gamma^{5}$ & $\frac{1}{2}\Gamma^{1}\Gamma^{6}$   & $\frac{1}{2}\Gamma^{1}\Gamma^{7}$ & $\frac{1}{2}\Gamma^{1}\Gamma^{8}$ & $s^{1 9}\equiv \frac{1}{2}\Gamma^{1}$ \\
\hline
\rule{0pt}{5mm} &  & $\frac{1}{2}\Gamma^{2}\Gamma^{3}$  & $\frac{1}{2}\Gamma^{2}\Gamma^{4}$ & $\frac{1}{2}\Gamma^{2}\Gamma^{5}$ & $\frac{1}{2}\Gamma^{2}\Gamma^{6}$ & $\frac{1}{2}\Gamma^{2}\Gamma^{7}$ & $\frac{1}{2}\Gamma^{2}\Gamma^{8}$ & $s^{2 9}\equiv \frac{1}{2}\Gamma^{2}$  \\
\hline
\rule{0pt}{5mm}  &  &  & $\frac{1}{2}\Gamma^{3}\Gamma^{4}$ & $\frac{1}{2}\Gamma^{3}\Gamma^{5}$ & $\frac{1}{2}\Gamma^{3}\Gamma^{6}$ & $\frac{1}{2}\Gamma^{3}\Gamma^{7}$ & $\frac{1}{2}\Gamma^{3}\Gamma^{8}$ & $s^{3 9}\equiv \frac{1}{2}\Gamma^{3}$ \\
\hline
\rule{0pt}{5mm} &  &   &  & $\frac{1}{2}\Gamma^{4}\Gamma^{5}$ & $\frac{1}{2}\Gamma^{4}\Gamma^{6}$ & $\frac{1}{2}\Gamma^{4}\Gamma^{7}$ & $\frac{1}{2}\Gamma^{4}\Gamma^{8}$ & $s^{4 9}\equiv \frac{1}{2}\Gamma^{4}$ \\
\hline
\rule{0pt}{5mm} &  &  &  &  & $\frac{1}{2}\Gamma^{5}\Gamma^{6}$ & $\frac{1}{2}\Gamma^{5}\Gamma^{7}$ & $\frac{1}{2}\Gamma^{5}\Gamma^{8}$ & $s^{5 9}\equiv \frac{1}{2}\Gamma^{5}$ \\
\hline
\rule{0pt}{5mm} &  &  &  &  &  &  $\frac{1}{2}\Gamma^{6}\Gamma^{7}$ &  $\frac{1}{2}\Gamma^{6}\Gamma^{8}$ & $s^{6 9}\equiv \frac{1}{2}\Gamma^{6}$ \\
\hline
\rule{0pt}{5mm} &  &  &  &  &  &  &  $\frac{1}{2}\Gamma^{7}\Gamma^{8}$ & $s^{7 9}\equiv \frac{1}{2}\Gamma^{7}$ \\
\hline
\rule{0pt}{5mm} &  &  &  &  &  &  &  &  $s^{8 9}\equiv \frac{1}{2}\Gamma^{8}$ \\
\hline
\end{tabular}
\label{Table_2}
\end{center}

The gamma matrices in Table 2 are taken from the set (\ref{Main-2}).

The dimension of the SO(m,n) algebra is given by $\frac{(m+n)(m+n-1)}{2}=\frac{10\cdot 9}{2}=45$. Therefore, here we deal with the representation of SO(1,9) algebra.

Table 2 demonstrates not only the explicit form of the generators (\ref{Spin-2}) but the commutation relations (\ref{Comm-2}) as well. Indeed, it is evident that generators with different indices commute between each other. Here again we have three independent sets of SU(2) generators ($s^{1}=s^{23},\,s^{2}=s^{31},\,s^{3}=s^{12}$), which commute between each other. Therefore, here similarly to the consideration in Sec. 4 the spin 3/2 Lorentz and Poincar\'e symmetries for the equation suggested in \cite{Sim-1,Sim-2} and \cite{Book} can be found.

\section{Briefly on application to symmetry analysis}\label{symmetry}

Consider only evident result that the 21 dimensional gamma matrix representation of the subalgebra SO(7) of the algebra SO(10), which is formed by the operators
\begin{equation}
\{s^{\breve{\mathrm{A}}\breve{\mathrm{B}}}\}=\{s^{\breve{\mathrm{A}}\breve{\mathrm{B}}}\equiv\frac{1}{4}[\Gamma^{\breve{\mathrm{A}}},\Gamma^{\breve{\mathrm{B}}}]\}, \quad \breve{\mathrm{A}},\breve{\mathrm{B}}=\overline{1,7},
\label{Solit}
\end{equation}
determines the pure matrix algebra of invariance of the Dirac equation in the Foldy--Wouthuysen representation (\ref{Foldy}). The complete set of the pure matrix symmetries of equation (\ref{Foldy}) is given by 21 elements of SO(7) plus three SU(2) operators ($s^{1}_{\mathrm{III}}=-\frac{1}{2}\Gamma^{8}\Gamma^{9}, \, s^{2}_{\mathrm{III}}=-\frac{1}{2}\Gamma^{9}\Gamma^{7}, \, s^{3}_{\mathrm{III}}=-\frac{1}{2}\Gamma^{7}\Gamma^{8}$). The corresponding symmetries of the Dirac equation can be found by inverse Foldy--Wouthuysen transformation \cite{Foldy} in the space of 8-component spinors. Note that in the Dirac representation the main part of these operators will not be pure matrix. The usefulness of the Foldy--Wouthuysen transformation is demonstrated recently in \cite{Silenko,Silenko-2} and in our above mentioned papers as well. 

The 84 dimesinal set of pure matrix operators can be found by multiplication of 21 elements of SO(7) by each element from the given here SU(2) set. However, it will be the overlapping algebra. Note that sometimes the overlapping algebra can be useful as well. We can recall the 31-dimensinal algebra C(1,3)$\oplus\varepsilon$C(1,3)$\oplus\varepsilon$, where $\varepsilon$ is the duality transformation (the transformation of Heaviside--Larmore--Rainich in the space of field-strengths of electromagnetic field). Such maximal first-order symmetry of the Maxwell equations in the terms of field strengths was found in \cite{Kri-3}. The usefulness was demonstrated e.g. in \cite{Fush} and in many papers of other authors, which, unfortunately, often have forgotten to refer on \cite{Kri-3}.   

\section{Brief conclusions}\label{conclusions}

The suggested gamma matrix representations of different algebras open new possibilities for the investigations of the field theory equations for the higher spins, especially for the spin s=3/2.

\section{Appendix 1}

Below the 64 elements of the gamma matrix representation of the $\textit{C}\ell^{\mathbb{R}}$(0,6) algebra are given. We have first six elements from (\ref{Main}), 15 elements as pairs of operators

\vspace{0,1in}

\begin{tabular}{ccccccccc}

\rule{0pt}{5mm} $\Gamma^{1}\Gamma^{2}$  & $\Gamma^{1}\Gamma^{3}$ & $\Gamma^{1}\Gamma^{4}$ & $\Gamma^{1}\Gamma^{5}$ & $\Gamma^{1}\Gamma^{6}$ \\

\rule{0pt}{5mm} & $\Gamma^{2}\Gamma^{3}$ & $\Gamma^{2}\Gamma^{4}$  & $\Gamma^{2}\Gamma^{5}$ & $\Gamma^{2}\Gamma^{6}$ \\

\rule{0pt}{5mm} &  & $\Gamma^{3}\Gamma^{4}$  & $\Gamma^{3}\Gamma^{5}$ & $\Gamma^{3}\Gamma^{6}$ \\

\rule{0pt}{5mm}  &  &  & $\Gamma^{4}\Gamma^{5}$ & $\Gamma^{4}\Gamma^{6}$ \\

\rule{0pt}{5mm} &  &   &  & $\Gamma^{5}\Gamma^{6}$, \\

\end{tabular}

\vspace{1,5in}

20 elements as operator triplets

\vspace{0,1in}

\begin{tabular}{cccccccccccccccc}

\rule{0pt}{5mm} $\Gamma^{1}\Gamma^{2}\Gamma^{3}$  & $\Gamma^{1}\Gamma^{2}\Gamma^{4}$ & $\Gamma^{1}\Gamma^{2}\Gamma^{5}$ & $\Gamma^{1}\Gamma^{2}\Gamma^{6}$ &  &  &  & $\Gamma^{2}\Gamma^{3}\Gamma^{4}$ & $\Gamma^{2}\Gamma^{3}\Gamma^{5}$ & $\Gamma^{2}\Gamma^{3}\Gamma^{6}$ &  &  &  & $\Gamma^{3}\Gamma^{4}\Gamma^{5}$ & $\Gamma^{3}\Gamma^{4}\Gamma^{6}$ \\

\rule{0pt}{5mm} & $\Gamma^{1}\Gamma^{3}\Gamma^{4}$ & $\Gamma^{1}\Gamma^{3}\Gamma^{5}$  & $\Gamma^{1}\Gamma^{3}\Gamma^{6}$ &  &  &  &  & $\Gamma^{2}\Gamma^{4}\Gamma^{5}$ & $\Gamma^{2}\Gamma^{4}\Gamma^{6}$ &  &  &  &  & $\Gamma^{3}\Gamma^{5}\Gamma^{6}$\\

\rule{0pt}{5mm} &  & $\Gamma^{1}\Gamma^{4}\Gamma^{5}$  & $\Gamma^{1}\Gamma^{4}\Gamma^{6}$ &  &  &  &  &  & $\Gamma^{2}\Gamma^{5}\Gamma^{6}$ \\

\rule{0pt}{5mm}  &  &  & $\Gamma^{1}\Gamma^{5}\Gamma^{6}$  &  &  &  &  &  &  &  &  &  &  & $\Gamma^{4}\Gamma^{5}\Gamma^{6}$, \\

\end{tabular}

\vspace{0,1in}

15 elements as products of four operators

\vspace{0,1in}

\begin{tabular}{ccccccccccccccccc}

\rule{0pt}{5mm} $\Gamma^{1}\Gamma^{2}\Gamma^{3}\Gamma^{4}$  & $\Gamma^{1}\Gamma^{2}\Gamma^{3}\Gamma^{5}$ & $\Gamma^{1}\Gamma^{2}\Gamma^{3}\Gamma^{6}$ & &  &  & $\Gamma^{1}\Gamma^{3}\Gamma^{4}\Gamma^{5}$ & $\Gamma^{1}\Gamma^{3}\Gamma^{4}\Gamma^{6}$ &  &  &  & $\Gamma^{1}\Gamma^{4}\Gamma^{5}\Gamma^{6}$ \\

\rule{0pt}{5mm} & $\Gamma^{1}\Gamma^{2}\Gamma^{4}\Gamma^{5}$ & $\Gamma^{1}\Gamma^{2}\Gamma^{4}\Gamma^{6}$  &  &  &  &  & $\Gamma^{1}\Gamma^{3}\Gamma^{5}\Gamma^{6}$ \\

\rule{0pt}{5mm} &  & $\Gamma^{1}\Gamma^{2}\Gamma^{5}\Gamma^{6}$ \\

\end{tabular}

\begin{tabular}{ccccccccccccccccc}

\rule{0pt}{5mm} $\Gamma^{2}\Gamma^{3}\Gamma^{4}\Gamma^{5}$  & $\Gamma^{2}\Gamma^{3}\Gamma^{4}\Gamma^{6}$  &  &  &  & $\Gamma^{2}\Gamma^{4}\Gamma^{5}\Gamma^{6}$  &  &  &  & $\Gamma^{3}\Gamma^{4}\Gamma^{5}\Gamma^{6}$ \\

\rule{0pt}{5mm} & $\Gamma^{2}\Gamma^{3}\Gamma^{5}\Gamma^{6}$,  \\

\end{tabular}

\vspace{0,1in}

6 elements as products of five operators

\vspace{0,1in}

\begin{tabular}{ccccccccccccccccc}

\rule{0pt}{5mm} $\Gamma^{1}\Gamma^{2}\Gamma^{3}\Gamma^{4}\Gamma^{5}$  & $\Gamma^{1}\Gamma^{2}\Gamma^{3}\Gamma^{4}\Gamma^{6}$  &  &  &  & $\Gamma^{1}\Gamma^{2}\Gamma^{4}\Gamma^{5}\Gamma^{6}$  &  &  &  & $\Gamma^{1}\Gamma^{3}\Gamma^{4}\Gamma^{5}\Gamma^{6}$ &  &  &  & $\Gamma^{2}\Gamma^{3}\Gamma^{4}\Gamma^{5}\Gamma^{6}$ \\

\rule{0pt}{5mm} & $\Gamma^{1}\Gamma^{2}\Gamma^{3}\Gamma^{5}\Gamma^{6}$, \\

\end{tabular}

\vspace{0,1in}

one product of six operators $\Gamma^{1}\Gamma^{2}\Gamma^{3}\Gamma^{4}\Gamma^{5}\Gamma^{6}$ and $\mathrm{I}_{8}$ as unit element of the algebra.

\vspace{0,1in}

\section{Appendix 2}

The 256 elements of the gamma matrix representation of the $\textit{C}\ell^{\mathbb{R}}$(0,8) algebra can be recalculated similarly to the method presented in Appendix 1. We have eight independent elements from (\ref{Main}), 28 elements as pairs of operators, 56 elements as operator triplets, 70 products of four operators, 56 elements as products of five operators, 28 products of six operators, 8 products of seven operators, one product of eight operators $\Gamma^{1}\Gamma^{2}\Gamma^{3}\Gamma^{4}\Gamma^{5}\Gamma^{6}\Gamma^{7}\Gamma^{8}$ and $\mathrm{I}_{8}$ as unit element of the algebra.



\begin{thebibliography}{9}

\bibitem{Real-1}
V.M. Simulik, I.Yu. Krivsky. Bosonic symmetries of the Dirac equation, Phys. Lett. A., vol. 375, is. 25, 2479--2483 (2011).

\bibitem{Real-2}
V.M. Simulik, I.Yu. Krivsky, I.L. Lamer. Some statistical aspects of the spinor field Fermi--Bose duality, Cond. Matt. Phys., vol. 15, is. 4, 43101(1--10) (2012).

\bibitem{Real-3}
V.M. Simulik, I.Yu. Krivsky, I.L. Lamer. Bosonic symmetries, solutions and conservation laws for the Dirac equation with nonzero mass, Ukr. J. Phys., vol. 58, is. 6, 523--533 (2013).

\bibitem{Real-4}
V.M. Simulik. On the gamma matrix representations of SO(8) and Clifford Algebras, Adv. Appl. Clifford Algebras, vol. 28, is. 5, 93(1--15) (2018).

\bibitem{Book}
V. Simulik. Relativistic quantum mechanics and field theory of arbitrary spin, Nova Science, New York, 2020, 343 p.

\bibitem{Loun}
P. Lounesto. Clifford algebras and spinors, 2-nd edition, Cambridge University Press, Cambridge, 2001, 338 p.

\bibitem{Shir}
D.S. Shirokov. Clifford  algebras and spinors. Steklov Mathematical Institute, Moscow, 2011, 173 p. (in russian).

\bibitem{Eli}
J. Elliott, P. Dawber. Symmetry in Physics, vol.1, Macmillian Press, London, 1979, 366 p.

\bibitem{Wyb}
B. Wybourne. Classical groups for physicists, John Wiley and sons, New York, 1974, 415 p.

\bibitem{Mex}
V.M. Simulik, I.Yu. Krivsky. Clifford algebra in classical electrodynamical hydrogen atom model, Adv. Appl. Cliff. Algebras, vol. 7, is. 1, 25--34 (1997).

\bibitem{Elect}
What is the electron? Edit. V.M. Simulik, Apeiron, Montreal, 2005, 282 p.

\bibitem{Kri-1}
I.Yu. Krivsky, V.M. Simulik. The Dirac equation and spin 1 representations, a connection with symmetries of the Maxwell equations, Theor. Math. Phys., vol. 90, is. 3, 265--276 (1992).

\bibitem{Kri-2}
I.Yu. Krivsky, R.R. Lompay, V.M. Simulik. Symmetries of the complex Dirac--K$\ddot{\mathrm{a}}$hler equation, Theor. Math. Phys., vol. 143, is. 1, 541--558 (2005).

\bibitem{Sim-1}
V.M. Simulik. Derivation of the Dirac and Dirac-like equations of arbitrary spin from the corresponding relativistic canonical quantum mechanics, Ukr. J. Phys, vol. 60, is. 10, 985--1006 (2015).

\bibitem{Sim-2}
V.M. Simulik. Link between the relativistic canonical quantum mechanics of arbitrary spin and the corresponding field theory, J. Phys: Conf. Ser., vol. 670, 012047(1--16) (2016).

\bibitem{Good}
R.H. Good, Jr. Properties of the Dirac matrices, Rev. Mod. Phys., vol. 27, is. 2, 187--211 (1955).

\bibitem{Foldy}
 L.L. Foldy, S.A. Wouthuysen. On the Dirac theory of spin 1/2 particles and its non-relativistic limit, Phys. Rev., vol.78, is.1, 29–36 (1950).

\bibitem{Silenko}
A.J. Silenko, Exact form of the exponential Foldy--Wouthuysen transformation operator for an arbitrary-spin particle, Phys. Rev. A., vol.94, is. 3, 032104(1--6) (2016).

\bibitem{Silenko-2}
L. Zou, P. Zhang, A.J. Silenko, Position and spin in relativistic quantum mechanics, Phys. Rev. A., vol. 101, is. 3, 032117(1--19) (2020).

\bibitem{Kri-3}
I.Yu. Krivsky, V.M. Simulik. Lagrangian for the electromagnetic field in the terms of field strengths and the conservation laws, Ukr. J. Phys., vol. 30, is. 10, 1457--1459 (1985) (in russian).

\bibitem{Fush}
W.I. Fushchich, I.Yu. Krivsky, V.M. Simulik. On vector and pseudovector Lagrangians for electromagnetic field, Nuovo Cim. B., vol. 103, is. 4, 423--429 (1989).



\end{thebibliography}



\end{document}